\documentclass[aps,twocolumn,pre]{revtex4}

\usepackage{graphicx}
\usepackage{amsmath}

\usepackage{epsfig}

\newcommand{\e}{\epsilon}
\newcommand{\si}{\ket{\Psi(t)}}

\newcommand{\bra}[1]{\left\langle #1\right|}
\newcommand{\ket}[1]{\left|#1\right\rangle}

\newcommand{\bracket}[3]{\left\langle #1\left|#2\right|#3\right\rangle}
\newcommand{\cells}[4]{\left(#1 \otimes #2 \right) \otimes \left( #3 \otimes #4 \right)}

\newcommand{\av}[1]{\left\langle #1\right\rangle}
\def\cal#1{\mathcal{#1}}

\def\beq{\begin{equation}}
\def\eeq{\end{equation}}
\def\bea{\begin{eqnarray}}
\def\eea{\end{eqnarray}}

\def\buu{\bra{11}}
\def\bud{\bra{10}}
\def\bdu{\bra{01}}
\def\bdd{\bra{00}}
\def\bu{\bra{1'}}
\def\bd{\bra{0'}}
\def\kuu{\ket{11}}
\def\kud{\ket{10}}
\def\kdu{\ket{01}}
\def\kdd{\ket{00}}
\def\ku{\ket{1'}}
\def\kd{\ket{0'}}

\def\sf{\hat{\sigma}}

\begin{document}
\title{Two-stage coarsening mechanism in a kinetically constrained model of an attractive colloid}

\author{Stephen Whitelam$^1$} 
\author{Phillip L. Geissler$^{1,2}$}
\affiliation{$^1$~Department of Chemistry, University of California, and Physical Biosciences Division $^2$and Materials Sciences Division, Lawrence Berkeley National Laboratory, Berkeley, CA 94720}
\date{\today}
\begin{abstract}
We study an attractive version of the East model using the real-space renormalization group (RG) introduced by Stella et al. The former is a kinetically constrained model with an Ising-like interaction between excitations, and shows striking agreement with the phenomonology of attractive colloidal systems. We find that the RG predicts {\em two} nonuniversal dynamic exponents, which suggests that in the out-of-equilibrium regime the model coarsens via a two-stage mechanism. We explain this mechanism physically, and verify this prediction numerically. In addition, we predict that the characteristic relaxation time of the model is a non-monotonic function of attraction strength, again in agreement with numerical results.
\end{abstract}
\maketitle
\pagestyle{plain}
\setcounter{page}{1}
\setcounter{equation}{0}

\section{Introduction}

In this paper we study the attractive East model introduced by Geissler and Reichman~\cite{PhillnDave}, a kinetically constrained model designed to capture the physics of colloidal suspensions with short-ranged attractions~\cite{expts}. Kinetically constrained models (KCMs)~\cite{Fredrickson-Andersen,Palmer-et-al,Jackle-Eisinger,Kob-Andersen}
are systems in which certain trajectories between configurations are
suppressed~\cite{Whitelam-Garrahan}.  As a result, they possess
interesting slow dynamics
\cite{Schulz-Trimper,Sollich-Evans,Crisanti-et-al,Chung-et-al,Toninelli-et-al,Garrahan-Chandler,Berthier-Garrahan}. Simple KCMs, such as the facilitated spin models introduced by
Fredrickson and Andersen~\cite{Fredrickson-Andersen} (hereafter the FA
model) and J\"{a}ckle and Eisinger~\cite{Jackle-Eisinger} (hereafter
the East model) display the slow, cooperative relaxation
characteristic of supercooled liquids near the glass
transition~\cite{Garrahan-Chandler,Berthier-Garrahan}.  For a
review of the glassy dynamics of KCMs see~\cite{Ritort-Sollich}.

The attractive East model possesses in addition to a kinetic constraint a static attraction between facilitating excitations. It was shown numerically~\cite{PhillnDave} that this model captures some key features of the phenomenology of attractive colloids, such as a re-entrant (non-monotonic) timescale associated with relaxation in equilibrium, and logarithmic relaxation near re-entrance~\cite{expts}.

In this paper we study the attractive East model using the real-space RG scheme of References~\cite{Stella, Hooyberghs}. Our reasons for doing so are twofold. First, we wish to determine analytically the nature of the model's out-of-equilibrium coarsening behaviour. We expect this behaviour to caricature that of a real colloid whose attraction strength is fixed but whose equilibrium free volume alters suddenly in response to a pressure change. Such experiments are needed to distingush between various scenarios of attractive glassiness that account for equilibrium behaviour. The equilibrium behaviour of the attractive East model, modeling a set of experiments in which free volume remains fixed but attraction strength varies, was explored in Reference~\cite{PhillnDave}. Second, we wish to test the utility of the real-space RG scheme, which has previously been applied to quantum spin systems~\cite{Stella}, reaction-diffusion systems~\cite{Hooyberghs} and simple kinetically constrained models~\cite{RSRG}.

Our key results, which we present in Section~\ref{sec:study1}, are as follows. We show that the first-order RG scheme predicts {\em two} nonuniversal (parameter-dependent) dynamic exponents, which suggests that in the out-of-equilibrium regime the model coarsens via a two-stage mechanism. On short wavelengths, relaxation is approximately diffusive; on longer wavelengths, relaxation is dramatically sub-diffusive. We explain this mechanism in terms of a competition between the static attraction and the kinetic constraint, and verify this prediction numerically, as illustrated in Figure~\ref{fig:domains}. In addition, using an uncontrolled second-order adaptation of the RG scheme, we show in Section~\ref{sec:study2} that the characteristic relaxation time of the model is a non-monotonic function of attraction strength, in agreement with the numerical results of Ref~\cite{PhillnDave}. 

This paper is organised as follows. In Section \ref{sec:def} we define the attractive East model, and for convenience cast its evolution operator in the guise of a quantum spin model. In Section \ref{sec:rg} we explain the RG scheme of References~\cite{Stella, Hooyberghs}. In Section \ref{sec:east} we apply this RG to the ordinary East model, emphasising the physical interpretation of the scheme, and in Section \ref{sec:aeast} we do the same to its attractive counterpart. We thus demonstate that the formalism can, with little modification, treat both models. Those readers interested principally in the results of our study should focus on Sections \ref{sec:study1} and \ref{sec:study2}, in which we present our findings, to first- and second-order respectively, summarised above. We conclude in Section \ref{sec:conc}.

\section{The model}
\label{sec:def}
The attractive East model introduced in Reference~\cite{PhillnDave} is a chain of $N$ binary occupancies $n_i=0,1$ with periodic boundary conditions, an interacting reduced Hamiltonian
\beq
\label{hamil}
{\cal H}=-\e \sum_{i=1}^N n_i n_{i+1}+h(\e,c) \sum_{i=1}^N n_i,
\eeq
where $\e>0$, and the kinetic constraint of the East model~\cite{Jackle-Eisinger}. This constraint dictates that a cell $i$ may change state only if it possesses a left neighbour in the excited state $(n_{i-1}=1)$. We consider these occupancies to represent immobile $(n_i = 0)$ or mobile $(n_i = 1)$ regions of the colloid, defined over a suitable microscopic coarse-graining time~\cite{PhillnDave,Garrahan-Chandler}. We define the equilibrium concentration of mobile regions as $c$, and we shall refer to occupancies $n_i=1$ as `excitations'.

The attraction $\e$ induces correlations between excitations. It should not be taken to represent a literal attraction between mobile regions of a real colloid, but instead to describe the tendency of colloidal {\em particles} to attract one another~\cite{Phill-2}. These attractions may be mediated by dissolving a polymer in the colloid, or by changing the concentration of a dissolved salt solution. We regard the field $h(\e,c)$ as an auxiliary variable that depends on $c$ and $\e$ in such a way that $c$ is unchanged by varying $\e$: in other words, varying the strength of the inter-molecular attraction does not change the system's free volume. The required adjustment of field with attraction strength follows from the standard result for the magnetization of an Ising model in $d=1$~\cite{Chandler},
\beq
\label{conc-1}
h(\e,c) = \e + 2 \sinh^{-1} \left( \frac{1-2 c}{2 \sqrt{c(1-c)}} e^{-\e/2} \right).
\eeq
The dynamics of the model is governed by a master equation
\begin{equation}
\label{master}
\partial_t P(n,t)=-\sum_i w(n_i)
P(n,t) + \sum_i w(1-n_i) P(n',t),
\end{equation}
where $P(n,t)$ is the probability that the system has
configuration $n \equiv \{n_1,
\dots,n_i,\dots,n_N\}$ at time $t$ ($n'$ is the
configuration $n$ with occupancy $n_i \to 1-n_i$), and
$w(n_i)\equiv w(n_i,\{n_j\})$ is the probability per
unit time that $n_i$ will change state. The $\{n_j\}$ are the
nearest neighbours of $i$.

We now pass to a quantum formalism~\cite{Siggia} in the manner described in \cite{RSRG}. Equation~(\ref{master}) is recast as the Euclidean Schr\"odinger equation 
\beq
\label{schrod}
\partial_t \si=-\sum_i {\cal L}_i \, \si,
\eeq
where the Liouvillian ${\cal L}_i$ is
\bea
\label{liouv}
{\cal L}_i= {\cal C}_i(\{s_j\})
\left\{ e^{ -\frac{1}{2} \left( {\cal H}(n')-{\cal H}(n) \right)} -e^{-\frac{1}{2} \left( {\cal H}(n)-{\cal H}(n') \right)} \sf_i \right\},
\eea
and $\sf_i$ is the spin-flip operator defined via $\sf_i f(n) \equiv f(n') \sf_i$. The state vector $\si$ can be written as
\beq
\si = \sum_{\{n \}} P(n,t) \ket{n},
\eeq
with $\ket{n} \equiv \prod_{i=1}^N \otimes \ket{n_i}$. The kinetic constraint is ${\cal C}_i(\{n_{j}\})=
 n_{i-1}$, and so suppresses the dynamics of cell $i$ if cell $i-1$ is unexcited.

We can write the Liouvillian (\ref{liouv}) as~\cite{RSRG}
\bea
\label{zero}
{\cal L}_i=1_1 \otimes \cdots \otimes n_{i-1} \otimes \ell_{i,(2)} \otimes n_{i+1}\otimes 1_{i+2} \otimes \cdots \otimes 1_N \nonumber \\
+1_1 \otimes \cdots \otimes n_{i-1} \otimes \ell_{i,(1)} \otimes v_{i+1}\otimes 1_{i+2} \otimes \cdots \otimes 1_N,
\eea
where $v_i \equiv 1-n_i$ is the vacancy operator. The full Liouvillian is composed of a sum of terms like those in (\ref{zero}), with one factor for each site of the lattice. All factors in (\ref{zero}) are $2 \times 2$ matrices. 

We can write the Liouvillian schematically as
\beq
\label{one}
{\cal L} = n \otimes \ell_2 \otimes n + n \otimes \ell_1 \otimes v,
\eeq
where we have dropped site labels for brevity. Using the representation 
\beq
\sf_i=\left(\begin{array}{cc}0&1\\1&0 \end{array} \right), \quad
n_i=\left(\begin{array}{cc}1&0\\0&0 \end{array} \right),
\eeq
we can write the single-site Liouvillians as
\beq
\label{two}
\ell_k \equiv \left( \begin{array}{cc} e^{h/2 -k \e/2} & -e^{-h/2+k \e/2} \\
     -e^{h/2 -k \e/2}& e^{-h/2+k \e/2} \end{array} \right),
\eeq
where the excitation number $k=1,2$. The off-diagonal elements of $\ell_k$ control the rates for the flipping processes: if $n_i$ has two neighbouring excitations ($k=2$) it will excite with rate $e^{-h/2+\e}$, and de-excite with rate $ e^{h/2-\e}$ [note the global minus sign in Equation (\ref{schrod})]. If only its left neighbour is excited ($k=1$), $n_i$ will excite with rate $e^{-h/2+\e/2}$ and de-excite with rate $e^{h/2-\e/2}$. Since $h-\e$ is a monotonically decreasing function of $\e$~\cite{PhillnDave}, the attraction $\e$ encourages excitations to congregate. 

The diagonal elements of $\ell_k$ are such that the sum of elements in each column is zero, a property that is required by any probability-conserving stochastic process. We shall call this property {\em stochasticity}~\cite{Hooyberghs}, and we shall require that it be preserved under renormalization.

\section{Real-space renormalization group}
\label{sec:rg}
We shall study the attractive East model using a real-space RG developed in the 1980s, and since applied to quantum spin models~\cite{Stella}, reaction-diffusion systems~\cite{Hooyberghs} and nonequilibrium exclusion models~\cite{Hannay}. The procedure is as follows. First, one partitions the lattice into blocks of $b$ spins; we shall take $b=2$. Then one splits the Liouvillian ${\cal L}$ into a reference, intra-block piece ${\cal L}_0$, and an inter-block interaction $V$ in which an expansion will be made. By performing a suitable coarse-graining from the original lattice to a renormalized block-spin lattice, one obtains a coarse-grained or renormalized Liouvillian from which one may infer the scaling properties for the model in question. In Ref.~\cite{RSRG} we showed that this scheme yields the low-temperature critical behaviour of some simple kinetically constrained models in one dimension. 

Because the attractive East model possesses 3-site interaction terms we must go to ${\cal O}(V^2)$ in the perturbation expansion. We shall outline the procedure; see \cite{Stella} for more details.

We first split the Liouvillian into a reference piece and a perturbation,
\beq
{\cal L} = {\cal L}_0 + V.
\eeq
This splitting is arbitrary, and we shall choose the reference piece ${\cal L}_0$ to be the simplest term that we expect to capture the physics of the model under study, subject to the requirement that it consist only of intra-block operators. The interaction $V$, which constitutes the remainder of the Liouvillian, includes in addition inter-block interactions. For example, ${\cal L}_0$ might look like
\beq
{\cal L}_0 = \left (n \otimes \ell \right) \otimes \left(1 \otimes 1 \right),
\eeq
where the brackets indicate the way in which we choose to partition sites into renormalized cells. The inter-block interaction $V$ might then look like
\beq
V = \left(1 \otimes n \right) \otimes \left(\ell \otimes 1 \right),
\eeq
where now the nontrivial operators $n$ and $\ell$ are split between the notional renormalized cells. 

Next, we introduce the right eigenstates $\ket{\phi_i} = \left\{  \ket{g_j},\ket{e_l} \right\}$ of the reference Liouvillian, 
\beq
{\cal L}_0 \ket{\phi_i} = E_{\phi_i} \ket{\phi_i}, 
\eeq
where $\ket{\phi_i}$ includes both ground states $\ket{g_j}$ with $E_{g_j}=0$, and excited states $\ket{e_l}$ with $E_{e_l} \neq 0$. We shall find that the reference state we choose for the attractive East model has three ground states $(j=1,2,3)$ and one excited state $(l=1)$. Ground states describe slow processes, and excited states describe fast processes. Note that here and subsequently the excited states to which we refer are excited states of Liouvillians (evolution operators), rather than Hamiltonians (energy functions). The original RG scheme~\cite{Stella} indeed involved Hamiltonians, whereas the subsequent extension of the scheme to nonequilibrium processes~\cite{Hooyberghs}, used in this paper, exchanged Hamiltonians for Liouvillians by exploiting the similarity between equations such as (\ref{schrod}), and bone fide Schr\"{o}dinger equations.

We require also the left eigenstates of the reference Liouvillian,
\beq
\bra{\phi_i} {\cal L}_0 = \bra{\phi_i} E_{\phi_i},
\eeq 
which are generally distinct from the right eigenstates for non-Hermitian Liouvillians.

The renormalization is executed by projecting the Liouvillian onto an arbitrarily-chosen coarse-graining subspace ${\cal S}$ of the reference piece, corresponding to its ground- or low-lying excited states~\cite{Hooyberghs, Stella}. What remains is an evolution operator for slow, coarse-grained dynamical processes; whether this captures successfully the physics of the model depends on how valid are the arbitrary choices one makes, both of the reference piece and of the coarse-graining subspace.

We shall construct the coarse-graining subspace ${\cal S}$ from only the ground states of the reference Liouvillian, in which case we have~\cite{Hooyberghs,Stella}
\bea
{\cal L}_0'(i,j) &=& 0;  \\
\label{first}
{\cal L}_1'(i,j)&=&\bracket{G_i}{V}{G_j}; \\
\label{second}
{\cal L}_2'(i,j)&=&- \sum_k \bracket{G_i}{V}{\phi_k} \frac{1}{E_{\phi_k}}\bracket{\phi_k}{V}{G_j}, 
\eea
where the $\ket{G_{\alpha}}$ (and their left counterparts) are linear combinations of the ground-state vectors of ${\cal L}_0$. The subscripts on the matrix elements ${\cal L}(i,j)$ refer to the order of the perturbation series. The second order result (\ref{second}) is required only if the Liouvillian of the model under study possesses $n$-site interaction terms, with $n>2$, as is the case for the attractive East model. For the ordinary East model, which contains only two-site interactions, the first order term (\ref{first}) is sufficient. The sum in (\ref{second}) runs over all eigenstates of ${\cal L}_0$ not assigned to the subspace ${\cal S}$. We shall call this the complementary subspace, $\bar{ {\cal S}}$.

The factor of $E_{\phi_k}^{-1}$ in (\ref{second}) weights the relative contributions of excited states to the renormalization procedure. States with small eigenvalues, those corresponding to slower processes, are accorded more importance as a consequence of this factor.

The excited states divide into two classes: those containing single-cell excitations, and those containing two-cell excitations. Higher-order contributions vanish by virtue of the orthogonality between distinct eigenvectors, and the fact that the model we study contains at most three-site, or two-cell interactions. To second order we acquire contributions from one-cell and two-cell excitations.

Note that according to the RG prescription of Stella et al., the sum in the second-order result (\ref{second}) runs over {\em all} eigenstates of the reference  Liouvillian ${\cal L}_0$ not assigned to the subspace ${\cal S}$. In our treatment of the attractive East model we shall arbitrarily discard one of the right ground eigenstates of ${\cal L}_0$, $\ket{g_3}$, and we shall explain in detail in the following section our physical reasons for doing so. To first order in the RG scheme this omission introduces no mathematical complications. However, to second-order, the sum in Equation (\ref{second}) becomes ill-defined. We find that ${\cal L}_0$ admits one eigenstate with nonzero eigenvalue, and we can incorporate this into the sum without difficulty. However, by omitting $\ket{g_3}$ from the coarse-graining subspace ${\cal S}$, and thereby assigning it to the complementary subspace $\bar{{\cal S}}$, we are required by the RG to sum over this state in (\ref{second}). Two issues arise. The first is that while $\ket{g_3}$ is included in the subspace $\bar{ {\cal S}}$, we assign its left counterpart $\bra{g_3}$ to the coarse-graining subspace ${\cal S}$ (we choose to do this in order to ensure probability conservation). Thus the projection of $\bra{g_3}$ in the complementary subspace $\bar{ {\cal S}}$ is the zero vector. The second issue is that the eigenvalue associated with the omitted state is zero, and so the denominator $E_{g_3}$ vanishes. As a consequence of these two points, the relevant term in Equation (\ref{second}) contains a factor $0/0$, and so is undefined. Since we cannot treat this term in any meaningful way, we choose to discard it. It should therefore be borne in mind that at second order our treatment is entirely uncontrolled in the mathematical sense. Nonetheless we shall present our results obtained in this manner, which appear to reproduce the numerically-observed~\cite{PhillnDave} non-monotonic behaviour of the attractive East model's charateristic relaxation time.  

It might be possible to justify our treatment to second order as an embedding onto a three-state coarse-graining subspace ${\cal S}_3$ (spanning all of the ground space), followed by a projection onto the original two-dimensional space ${\cal S}$~\cite{Hannay, ref1}. This procedure would still be uncontrolled, however. An alternative means of regularizing the troublesome term in Equation (\ref{second}) might be to modify the physics of the model in a controllable way. The physical root of the mathematical problem discussed above is the (total) asymmetry of the attractive East model's dynamics. One could imagine replacing the attractive East model with a model interpolating between it and an attractive version of the FA model, similar to the interpolation model of Buhot and Garrahan~\cite{Buhot-Garrahan}. The RG properties of the latter were studied in~\cite{RSRG}, where it was shown that the ground-state vector $\ket{g_3}$ acquires a nonzero eigenvalue of ${\cal O}(r)$. Here $r$ is an asymmetry parameter equal to zero in the East model limit. By taking the attractive East model limit of the attractive interpolation model in a suitable way, it might be possible to regularize the $0/0$ pathology plaguing the current treatment.

In the following section we describe how this formalism may be used to treat the ordinary East model, emphasising the physical interpretation of the procedure. In section \ref{sec:aeast} we apply this formalism to the attractive East model.

\section{RG for the East model}
\label{sec:east}
In this section we shall outline the application of the RG procedure described above to the East model~\cite{RSRG}. We show that it is possible to partially define the scheme by imposing normalization and probability conservation. To complete the definition we arbitrarily discard part of the ground-state subspace of the reference Liouvillian, in what follows we justify this choice physically by considering in detail the peculiar hierarchical dynamics of the East model. This allows us to show that the RG for the attractive East model, which we discuss in the following section, can be determined, to first order, by straightforward generalization of the result for its noninteracting counterpart.

The Liouvillian of the East model is
\beq
\label{east}
{\cal L} = n \otimes \ell,
\eeq
where 
\beq
\label{two}
\ell \equiv \left( \begin{array}{cc} 1-c & -c \\
     c-1& c \end{array} \right),
\eeq
and $c$ is the excitation rate of a facilitated cell. Equation (\ref{east}) describes the dynamics of a cell facilitated only by a left-neighbouring excitation. The RG prescription (\ref{first}) may be written
\beq
\label{rg-prescription}
{\cal L}'(n') = T_1(n',n) \cdot {\cal L}(n)\cdot T_2(n,n'),
\end{equation}
where $T_1$ and $T_2$ are respectively $2^{N/2} \times 2^N$ and $2^N \times 2^{N/2}$-dimensional matrices built from the left $(T_1)$ and right $(T_2)$ ground-state eigenvectors of ${\cal L}_0$. The RG prescription is thus a mapping from a $2^{N}$-dimensional Hilbert space of `real' occupancies to a $2^{N/2}$-dimensional Hilbert space of renormalized occupancies, which we distinguish with primes.

As discussed in Ref.~\cite{RSRG}, we find that $T_1$ and $T_2$ have simple interpretations in terms of the dynamics of the model under study. The  projection matrix $T_1$ takes a `real' state $\{11,10,01,00\}$ and projects it onto a renormalized state. For the East model we choose that states with at least one excitation project onto a renormalized excitation. Thus $T_1=t_1^{\otimes N/2}$ where
\beq
\label{proj}
t_1=\ku \left\{\buu+\bud+\bdu\right\} + \kd \bdd
\eeq
Hence blocks of spins which can, within a single time step, facilitate neighbouring spins are deemed also to be facilitating in the coarse-grained sense. Note that if we were to renormalize an Ising model with no kinetic constraint, then we would require that flipping a spin in an unrenormalized configuration would result in the flipping of the coarse-grained spin. This is not the case for kinetically constrained models, for which there is no symmetry between excitations and vacancies.

The embedding matrix $T_2$ reconstitutes a real state from a coarse-grained state, and in a sense identifies those original states most important for the low temperature dynamics of the system. As discussed in~\cite{RSRG}, we build $T_2$ so as to respect both energetic and entropic effects. For the East model we choose $T_2=t_2^{\otimes N/2}$, where
\beq
\label{embed}
t_2= \frac{1}{1+ \lambda} \left\{ \lambda \kuu + \kud \right\} \bu + \kdd \bd.
\eeq
We have defined $\lambda \equiv c/(1-c)$. The structure of (\ref{embed}) can be partially determined by the right ground-state eigenvectors of ${\cal L}_0$, together with the requirements of normalization and conservation of probability~\cite{RSRG}, but may be simply motivated as follows. To find $t_2$, one adds to the term $\kdd \bd$ a sum of terms $(1+ \lambda)^{-1} \lambda^{n_1+n_2-1} \ket{n_1,n_2} \bu$, where $n_1 n_2$ is an unrenormalized configuration, and the power of $\lambda$ accounts for the energetic weighting of these states. Thus the state $11$ is penalized by a factor of $\lambda$, wheras the state $10$ receives no penalty.

According to this rule we should also include in~(\ref{embed}) a term $(1+ \lambda)^{-1} \kdu \bu$, but we choose to suppress this for the following reason. The East model possesses an hierarchical dynamics~\cite{Ritort-Sollich,Sollich-Evans}, whereby two excitations separated by a distance $d$ are relaxed by establishing a set of isolated excitations between them, at distances $d/2$, $3 d/4$ etc. We can incorporate this behaviour into our RG scheme by suppressing `frozen' configurations $01$ during embedding, which is permitted by the structure of the ground state eigenvectors of the East model reference Liouvillian. Under repeated application of this modified embedding operator we then see that the most important state in the dynamical sense for $d=4$ is
\beq
\label{state-1}
1011, \qquad \textnormal{weight } 1 \times c^2,
\eeq
and not, for instance,
\beq
\label{state-2}
1001, \qquad \textnormal{weight } 0 \times c.
\eeq
The latter state is favoured thermodynamically over the former, because it contains one fewer excitation, but the latter is suppressed entropically (the factor 0 multiplying $c^2$). To see this, note that in order to relax the rightmost excitation in state (\ref{state-2}) we must excite the second spin, followed by the third, and then finally we may relax the rightmost spin. To perform a similar relaxation for state (\ref{state-1}) we may simply relax the rightmost state. Thus in an approximate sense we see that our ad-hoc choice of $T_2$ respects the hierarchical dynamics of the East model. Note that for the FA model, which has symmetrical dynamical rules, we choose to make no such suppression. Indeed, none is required. The ground-state subspace of the FA model's reference Liouvillian is only doubly degenerate, and these two states are natural choices for the two-dimensional coarse-graining subspace ${\cal S}$. States (\ref{state-1}) and (\ref{state-2}) are then of similar importance to the dynamics.

The renormalization prescription can thus be thought of as a scheme that picks out those dynamical trajectories that most readily relax the system. As detailed in \cite{RSRG} we find for the East model the dynamic exponent $z=-1/(\ln c \ln 2)$, to leading order in $c$, and a marginally unstable critical point at $c=0$, at which the recursion relation for $\lambda = c/(1-c)$ satisfies $\lambda' = \lambda+ \lambda^2$. From these results and the relation between excitation concentration and temperature, $c=1/(1+e^{1/T})$, the scaling properties of the East model at low temperature follow. For example, based on the existence of a critical point we can write down a scaling relation for the density of excitations $n$ in the out-of-equilibrium regime,
\beq
\label{sc}
n(t,\lambda)=e^{-d \ell} \hat{n} \left(t(\ell) , \lambda(\ell) \right).
\eeq
Here $d=1$ is the physical dimension, $\ell=\ln b$ parameterizes the lattice rescaling parameter (now taken to be a real number), $\hat{n}$ is a dimensionless scaling function, and the arguments on the right hand side of (\ref{sc}) are the renormalized flowing time $t(\ell)= e^{-z \ell} \, t$ and temperature parameter $\lambda(\ell) = 1/\left(\lambda^{-1}-\ln(\ell/\ell_0)\right)$, respectively. Equation (\ref{sc}) relates a real system at long times and low densities (left hand side) to an effective, renormalized system at short times and high densities (right hand side). By imposing the matching condition $e^{-z \ell} \, t=1$ we find in the regime $0 \ll \ln t \ll 1/\left(T^2 \ln 2\right)$ the anomalous coarsening behaviour
\beq
n(t) \sim t^{-T \ln 2},
\eeq
which has been derived by other means and verified numerically~\cite{Ritort-Sollich}. 

\section{RG for the attractive East model}
\label{sec:aeast}

We now apply the formalism of Section 2 to the attractive East model, whose Liouvillian is 
\beq
\label{one-prime}
{\cal L} = n \otimes \ell_2 \otimes n + n \otimes \ell_1 \otimes v.
\eeq
In order to obtain a reference Liouvillian that contains only two-site interactions, we write the vacancy operator $v$ in the second term on the right hand side of (\ref{one-prime}) as $1-n$, to get
\beq
\label{one-p}
{\cal L} = n \otimes \ell_1 + n \otimes \Delta \otimes n,
\eeq
where $\Delta \equiv \ell_2-\ell_1$. We choose as our reference Liouvillian
\beq
{\cal L}_0 = n \otimes \ell_1. 
\eeq
The ground state embedding and projection operators are then similar to those for the ordinary East model. The projection operator $T_1$ is identical, which may be motivated in a physical way by considering that the definition of a facilitating spin in the attractive East model is identical to that in its ordinary counterpart. The embedding operator $T_2 = t_2^{\otimes N/2}$ now accounts for the attraction-modified Boltzmann weights of two-site configurations: the attraction `rewards' neighbouring excitations with an energy $-2 \e$ with respect to an excitation-vacancy pair. We enforce the same ad hoc suppression of the state $01$ as we did in the case of the ordinary East model. We thus get
\beq
\label{embed-a}
t_2=  \frac{1}{e^{\e-h} + 1}\left(e^{\e-h} \kuu +\kud \right) \bu + \kdd \bd,
\eeq
where $h=h(\e,c)$. In the limit $\e \to 0$ we recover Equation (\ref{embed}), with $\lambda \to e^{-h}$. For the second order result (\ref{second}) we require also the excited projection matrix $E_1 = e_1^{\otimes N/2}$ and embedding matrix $E_2 = e_2^{\otimes N/2}$, which are determined by
\beq
e_1 = \ku \left(\bud - e^{-\e+h} \buu \right),
\eeq
and
\beq
\label{e-two}
e_2 =\left(1+e^{h-\e} \right)^{-1} \left(-\kuu+\kud \right) \bu.
\eeq
The prefactor in (\ref{e-two}) ensures that $e_1 \cdot e_2 = \ket{1'} \bra{1'}$.

We now write our Liouvillian as ${\cal L} = {\cal L}_0+V$, where
\beq
{\cal L}_0= \cells{n}{\ell_1}{1}{1}
\eeq
and 
\beq
V= A \otimes B + C \otimes D,
\eeq 
with $A \equiv (1 \otimes n)$, $B \equiv (\ell_1 \otimes 1 + \Delta \otimes n)$, $C \equiv (n \otimes \Delta)$ and $ D \equiv (n \otimes 1)$. The brackets denote the blocking of the lattice into cells. 

We find under renormalization that to first order
\beq
\label{1st-order}
{\cal L}_1'= t_1   A   t_2 \otimes t_1   B   t_2.
\eeq
To second order, neglecting the pathological term discussed at the end of Section \ref{sec:rg}, we have
\bea
\label{2nd-order}
{\cal L}_2' = t_1   A   e_2   \cdot e_1   A   t_2 \otimes t_1   B   t_2 \cdot  t_1   B   t_2 + \nonumber \\
t_1   A   e_2 \cdot  e_1   C   t_2 \otimes t_1   B   t_2 \cdot  t_1   D   t_2 + \nonumber \\
t_1   A   t_2 \cdot  t_1   A   t_2 \otimes t_1   B   e_2  \cdot e_1   B   t_2 + \nonumber \\
\frac{1}{2} t_1   A   e_2 \cdot  e_1   A   t_2 \otimes t_1   B   e_2  \cdot e_1   B   t_2 + \nonumber \\
t_1   A   t_2  \otimes t_1   B   e_2 \cdot e_1 C t_2 \otimes  t_1 D t_2.
\eea
We can make sense of these results by noting that the combination $t_1   A   t_2 =n'$ is a renormalized number operator, and $t_1   B   t_2 =\ell'$ is a renormalized single-site Liouvillian. Hence the first order result (\ref{1st-order}) looks like a renormalized (ordinary) East model, ${\cal L}'= n' \otimes \ell'$. The second order 2-site terms, the first four lines of Equation (\ref{2nd-order}), also have the form `constraint $\otimes$ rate', or $n' \otimes \tilde{ \ell}'$. Note that the fourth term arises from a two-cell excitation, and so, by virtue of the factor of $E_{\phi_k}^{-1}$ in Equation (\ref{second}), enters with a factor of $\frac{1}{2}$. The final term in (\ref{2nd-order}) is a three-site interaction that allows us to write ${\cal L}_2$ in the form (\ref{one}).

In addition, to second order, we find two terms corresponding to operators not present in the original Liouvillian, namely
\bea
\label{prolif}
\tilde{{\cal L}}_2'= t_1   A   t_2  \otimes t_1   A   e_2 \cdot e_1 B t_2 \otimes  t_1 B t_2 + \nonumber \\
 t_1   A   t_2  \otimes t_1   B   e_2 \cdot e_1 A t_2 \otimes  t_1 B t_2.
\eea
For example, the second term in (\ref{prolif}) represents a non-local facilitation of the form $n_{i-1}' \otimes \cdots \otimes \ell_{i+1}'$. Hence to this order the renormalization procedure is not closed. We choose, arbitrarily, to ignore these terms, for two reasons. First, in order to treat them properly within the framework of the RG we would have to insert them into the unrenormalized Liouvillian, which, because of the terms' nonlocal nature we deem to be unphysical; and second, omitting these terms seems not to adversely affect our results.

\section{Results of our study to first order}
\label{sec:study1}
With the renormalized Liouvillian now in hand we can extract the scaling properties of the attractive East model in the limit of small defect concentrations and long times. The first order result (\ref{1st-order}) shows that in the long time-  and large length-scale limit the model behaves like an ordinary East model with renormalized parameters. This agrees with the analysis of~\cite{PhillnDave}, which concluded that the attractive East model behaves on long length scales as an ordinary East model, albeit with a rescaled typical width of excited domains. We shall use the second order RG result (\ref{2nd-order}) to infer the `flow' of the model towards this `renormalized' East-like behaviour. To extract the dynamic exponent, however, it is sufficient to use the first-order result, which describes the slowest dynamically relevant processes. We proceed as follows.

The dynamic exponent $z$, defined via $t' = b^{-z} t$, describes the rescaling of time as a consequence of rescaling space by a factor of $b^{-1}$. Hence we can determine $z$ by studying the ratio of renormalized to unrenormalized rates. We define the rates $\Gamma_{\alpha}$, $\alpha \in \{1,2,3,4\}$, for the four processes of the model as follows:
\begin{center}
\begin{tabular}{l|c|c}
\hline
$\alpha$&Process&$\Gamma_{\alpha}$\\
\hline
$1$&$111 \to 101$&$e^{h/2-\e}$\\
$2$&$110 \to 100$&$e^{h/2-\e/2}$\\
$3$&$101 \to 111$&$ e^{-h/2+\e}$\\
$4$&$100 \to 110$&$e^{-h/2+\e/2}$\\
\hline
\end{tabular}
\end{center}
Let us define the renormalized rates $\Gamma_{\alpha}^{(1)}$ extracted from the first order result (\ref{1st-order}). Then upon rearranging $\Gamma^{(1)} = b^{-z} \Gamma $ we have $z_{\alpha} = -\ln r_{\alpha}/\ln 2$, where $r_{\alpha} \equiv \Gamma_{\alpha}^{(1)}/\Gamma_{\alpha}$. The renormalized rates are cumbersome, and so we shall not display them explicitly.

\begin{figure}
\begin{center}
\psfig{ file=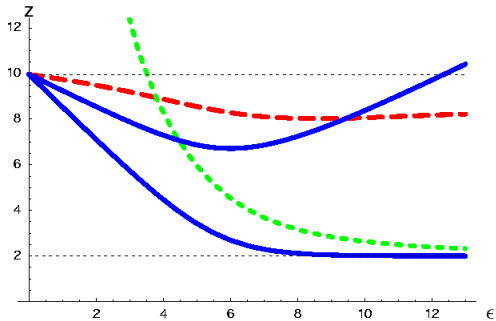,width=2.8in}
\end{center}
\caption{\label{fig:zed} (Colour online) Dynamic exponent $z$ versus attraction strength $\e$ for the attractive East model, at excitation concentration $c=10^{-3}$. The lower horizontal line marks the diffusive limit $z=2$; the upper horizontal line is the prediction for the East model in the absence of the attraction. The bold dashed lines are the phenomenological predictions $z_2$ (lower line, green) and $z_3$ (upper line, red). These we expect to be characteristic values for the dynamic exponent of the attractive East model when the domain sizes are respectively much less than and much greater than the attraction-induced lengthscale $a(\e,c)$. The bold blue lines are the prediction for $z$ from the first-order RG calculation. Both correctly reproduduce the $\e \to 0$ limit, and agree with our numerical results (see Figure~\ref{fig:domains}).}
\end{figure}

We find that, unlike for the ordinary East model, there is no common rescaling factor $z$. Instead we find four distinct $z_{\alpha}$, whose dependence upon $\e$ is dramatically different. Two values, $z_1$ and $z_4$, we discard as unphysical, being negative and tending asymptotially to unity, respectively. The other two we plot for excitation concentration $c=10^{-3}$ in Figure~\ref{fig:zed}. The larger value, $z_3$, is non-monotonic, and follows approximately the behaviour of the East-like exponent. The smaller value, $z_2$, tends to a diffusive value $2$ for large attraction strengths.

We can understand the nature of these two dynamic exponents from the following simple argument. The dynamic exponent characterizes the typical rate $\tau^{-1}$ at which a domain wall can move a given distance $\ell$, via $\tau \sim \ell^z$. Let us define a diffusion constant $D$, such that
\beq
\label{diff_const}
\ell^2 \equiv \av{(r(\tau)-r(0))^2} = 2D \tau,
\eeq
and $r(\tau)$ is the position of a given domain wall at time $\tau$. Replacing the exponent $2$ in Equation (\ref{diff_const}) by the more general value $z$, we can write
\beq
\label{gollum}
\tau \sim D^{-1} \ell^z.
\eeq
But if we consider domain wall drift as an `activated' process, one inhibited by an effective barrier of size $\Delta(\ell)$, then we can also write
\beq
\label{smeagol}
\tau \sim \Gamma_0^{-1} \exp(\Delta(\ell)),
\eeq
where $\Gamma_0 \sim D$ is the rate for attempting a barrier crossing. Equating (\ref{gollum}) and (\ref{smeagol}) and taking logarithms gives
\beq
z \sim \frac{\Delta(\ell)}{\ln \ell}.
\eeq
Note that the effective barrier $\Delta(\ell)$ accounts for the dynamics of the process under consideration. For example, for the ordinary East model the barrier $\Delta(\ell)$ grows logarithmically with $\ell$~\cite{Sollich-Evans}. We then obtain $z \sim 1/(T \ln 2)$ to leading order in $T$.

\begin{figure}
\begin{center}
\psfig{file=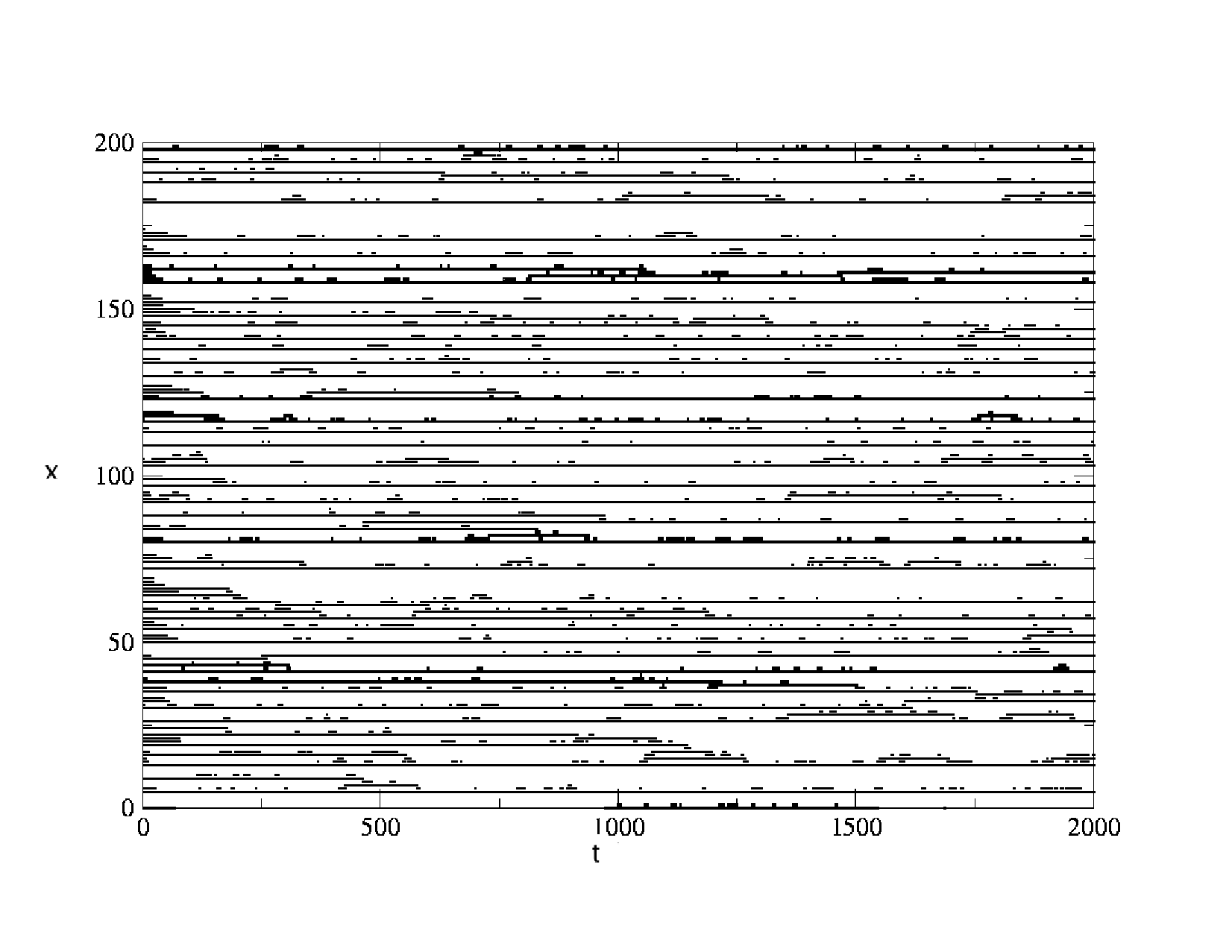,width=2.8in, angle=270}
\psfig{file=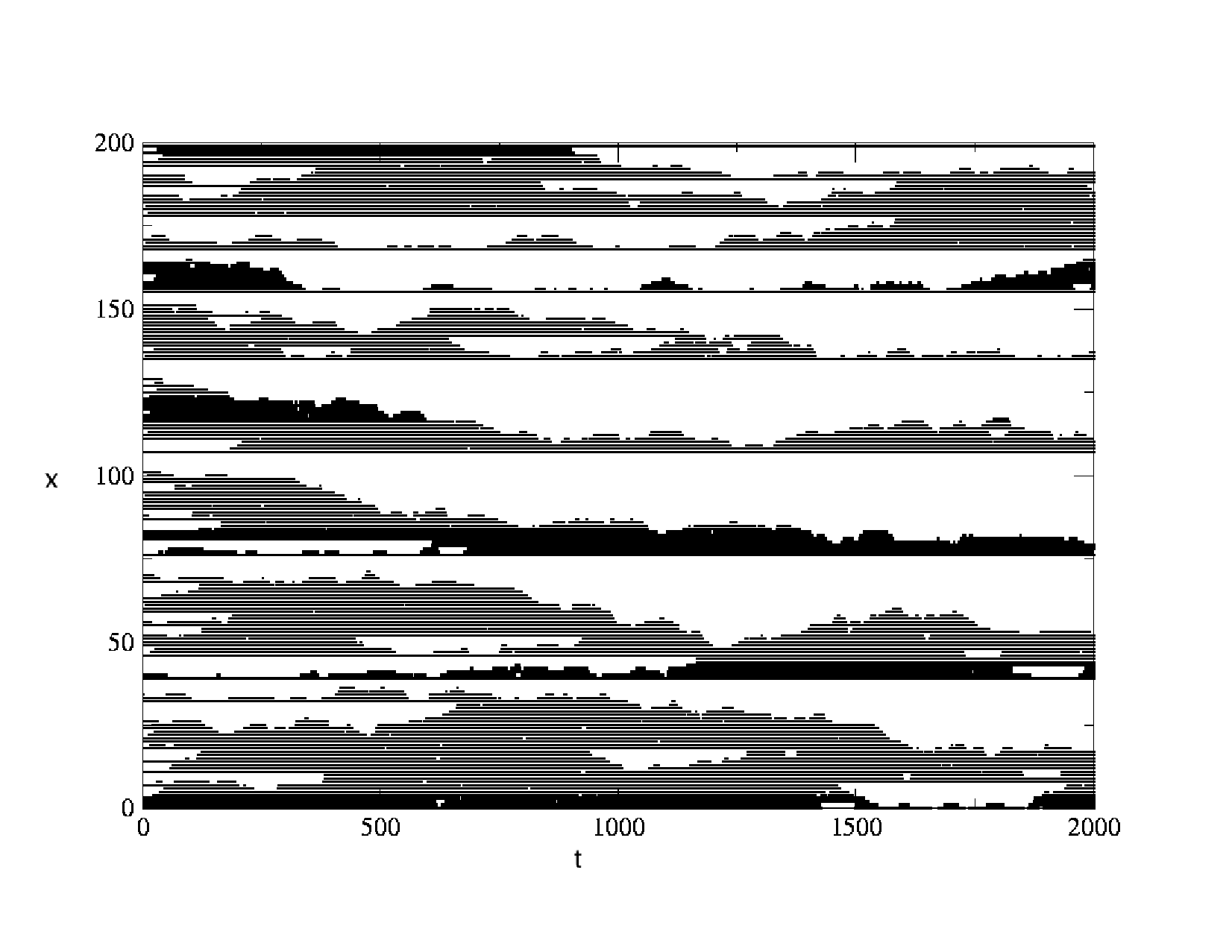,width=2.8in, angle=270}
\end{center}
\caption{\label{fig:trajectories}
An illustration of the finite thickness of excited (black) domain walls induced by the attraction $\e$. We show here nonequilibrium space-time trajectories for the attractive East model, starting from excitation concentration $n(0)=0.5$ for equilibrium excitation concentration $c=0.1$. We show the cases $\e=0$ (upper panel) and $\e=5$ (lower panel). Space runs along the vertical axis, time along the horizontal. Notice the emergence of a finite thickness of excited domains in the lower plot, for which the attraction-induced lengthscale $a \approx 24$ lattice sites. On lengthscales $\ell \leq a$ we expect essentially diffusive dynamics.}
\end{figure}

The two dynamic exponents we find in the case of the attractive East model arise because the attraction imposes a length scale below which the East-like hierarchical dynamics is suppressed. This follows from the fact that there exists in the attractive East model a characteristic width $a(c,\e)$ of excited (black)  domains~\cite{PhillnDave}. To see this, note that the energy of a configuration $10\cdots 01$ of length $a$ is $2h$. The energy of a similar length of chain with all the intermediate cells excited is $h+a(h-\e)$. These energies are equal when $a = h/(h-\e)$. Hence $a$, which increases as $\e$ increases, sets a length scale below which a 10 domain wall can readily move eastwards via a mechanism that excites contiguous cells. See Figure~\ref{fig:trajectories} for an illustration of this effect. This domain wall cannot move freely, however; the penalty for exciting successive cells to the east is, from $(1)$, $h-\e$. Thus domain wall motion on lengthscales $\ell \leq a$ should proceed principally by diffusion in a potential $(h-\e) \ell$. The largest barrier to be surmounted is therefore $(h-\e) a$, and so 
\beq
z_2 \sim \frac{(h-\e) a}{\ln a}.
\eeq   
We call this exponent $z_2$ as per the RG notation, and we expect it to dictate the dynamics of the system when the characteristic domain size $L(t)$ satisfies $L(t) \leq a$.

However, on lengthscales $L(t) > a$ we expect relaxation to proceed via East-like hierarchical dynamics, suitably rescaled to account for the attraction-induced length $a$. A characteristic value for the relevant dynamic exponent can be derived by considering the typical size of white domains in thermal equilibrium, $\av{L(t \to \infty )}= d(c,\e)$. This length may be found found from the partition function of the $d=1$ Ising model. Denoting by $\mu$ the largest eigenvalue of the transfer matrix
\beq
T_{n_i,n_{i+1}}=\left(\begin{array}{cc}
 1&e^{h/2}\\
e^{h/2}&e^{h-\e}\end{array}\right),
\eeq
we have that white domains of size $L$ occur with probability $P(L) =\exp(-L \ln \mu)$. Hence the typical white domain size in equilibrium is $d(c,\e)= 1/\ln \mu$. 
\begin{figure}
\begin{center}
\psfig{file=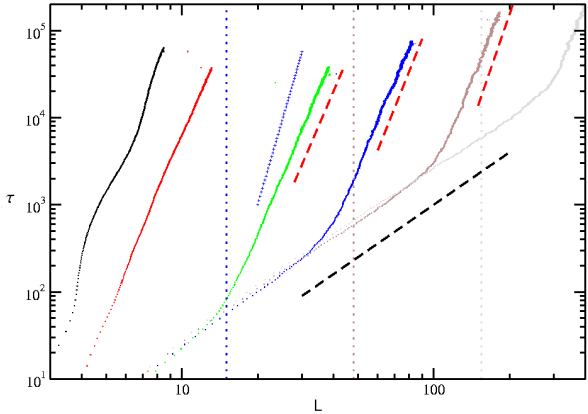,width=2.8in, angle=270}
\end{center}
\caption{\label{fig:domains} (Colour online) Continuous time Monte Carlo simulation data for the attractive East model with $c=10^{-3}$, showing elapsed time $\tau$ (vertical axis) versus mean white domain length $L$ (horizontal axis). The system size is $N=10^5$. We show the average of 10 runs for each attraction strength $\e=0,2,6,8,10,12$, from left to right. The two-stage coarsening mechanism is seen clearly in the three curves at largest $\e$, and the crossover occurs when the mean domain length exceeds approximately $a(c,\e)$. Note that the domain length distribution broadens as the model slows, and so the mean domain length becomes progressively less good as an estimate of the crossover. We show the value of $a$ as vertical dotted lines for $\e=8,10$ and $12$. We show also the RG predictions for short-wavelength coarsening at large $\e$,  $z = 2$ (heavy black dashed line), and for long-wavelength coarsening for $\e=6 \, (z \approx 6.72)$, $\e=8 \, (z \approx 7.24)$ and $\e=10 \, (z \approx 8.37)$ (heavy red dashed lines). We show for comparison next to the $\e=6$ line the ordinary East model low-temperature prediction, $z \approx 9.97$ (thin blue line), which clearly disagrees with the data.}
\end{figure}

The logarithmic barrier confronting domain walls moving a distance $d \gg a$ is then $\Delta(d) =h \ln(d/a)/\ln2$~\cite{PhillnDave}, and so
\beq
z_3 \sim \frac{ h \ln(d/a)}{\ln d \, \ln 2}.
\eeq
We show in Figure~\ref{fig:zed} the dynamic exponents calculated from both phenomenological and RG predictions. In Figure~\ref{fig:domains} we show numerical results for the attractive East model in the nonequilibrium regime. The coarsening mechanism is as we predict from RG and physical considerations: approximately diffusive relaxation on short wavelengths, crossing over to dramatically sub-diffusive behaviour on larger wavelengths.  Note that the value of $z$ we display for the ordinary East model is its low-temperature approximation. The crossover length (the position of the `kink' in the plots) is consistent with $\ell(t) \sim a$. For large values of $\e$ the distribution of white bubble lengths is very broad, and the crossover begins while the mean domain length is still appreciably less than $a$.

We verified also that diffusive coarsening persists throughout the nonequilibrium regime for systems such that $a>d$.

\section{Results of our study to second order}
\label{sec:study2}

While the first-order result is sufficient to determine the dynamic exponent of the model, to first order the renormalized Liouvillian (\ref{1st-order}) looks like that of an ordinary East model, in that the rate for creating an excitation is insensitive to the state of the right neighbour. However, the renormalized Liouvillian to second order, (\ref{2nd-order}), does account for the state of the right neighbour. We can infer from this result a characteristic timescale for relaxation, as well as a timescale for crossover from diffusive to sub-diffusive relaxation. It should be borne in mind that to second order our perturbation scheme is formally ill-defined. 

We can infer the values of the renormalized parameters via the matrix elements of the renormalized Liouvillian to second order, ${\cal L}'={\cal L}_1'+{\cal L}_2'$. In particular, we define the quantities
\beq
\label{params-1}
\alpha^{(2)} \equiv \frac{\Gamma_3^{(2)}}{\Gamma_1^{(2)}}; \quad 
\beta^{(2)} \equiv \frac{\Gamma_1^{(2)} \Gamma_4^{(2)} }{\Gamma_3^{(2)} \Gamma_2^{(2)}},
\eeq
whose unrenormalized counterparts are $\alpha = e^{2 \e -h(c,\e)}$ and $\beta = e^{-\e}$. From the latter two expressions we solve for $c$ and $\e$ as functions of $\alpha$ and $\beta$, obtaining $c=f(\alpha, \beta)$ and $\e = -\ln \beta$. The function $f$ is unwieldy, by virtue of the complicated dependence of $h(c,\e)$ upon its arguments, Equation (\ref{conc-1}), and so we shall not display it explicitly. Then we define renormalized parameters via $c' \equiv c^{(2)} = f(\alpha^{(2)},\beta^{(2)})$ and $\e' \equiv \e^{(2)} = - \ln \beta^{(2)}$.

In the absence of attraction the recursion relation for $c$ under renormalization reads $c=c/(1-c+c^2)$, encoding an unstable critical fixed point $c^{\star}=0$ and a stable full-lattice fixed point $c^{\star}=1$. We shall focus on the the regime $c \ll 1$, and shall avoid probing the large-$\e$ regime where possible static critical effects intrude. 

\begin{figure}
\begin{center}
\psfig{file=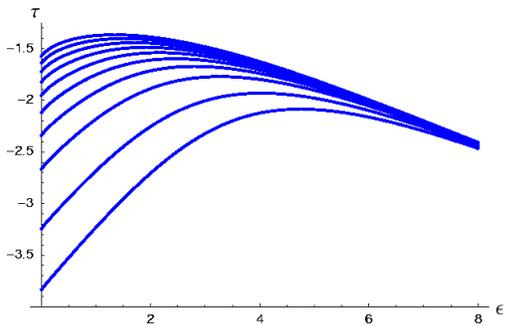,width=2.8in}
\end{center}
\caption{ \label{fig:timescale} (Colour online) Negative of the logarithm of the characteristic timescale $-\log_{10} \tau = \nu_{\parallel} \log_{10} c$ from the second-order RG result, for $c=0.01,0.02,0.04,0.06,0.08,0.10,0.12,0.14,0.16,0.18$ (bottom to top). The degree of non-monotonicity at each value of $c$ is in good agreement with the numerical results shown in Figure 3 of Reference ~\cite{PhillnDave}. The difference in absolute value at each concentration occurs because the RG result refers to a generic timescale $\tau$. For any given process $\tau_g$ there exists a basic rate $\Gamma(c, \e)$ such that $\tau_g = \Gamma(c,\e)\, c^{-\nu_{\parallel}}$, and hence the vertical offset  of $-\log_{10} \tau_g$ relative to the generic value $-\log_{10} \tau_g$ is $ -\log \Gamma(c, \e)$.}
\end{figure}

To extract a characteristic timescale we define the exponent $y_c$ via $c' = b^{y_c} c$. From standard RG arguments~\cite{Kadanoff} we then have that $\nu_{\perp} = 1/y_c$, where $\nu_{\perp}$ controls the divergence of the correlation length $\xi$ near the critical point $c=0$, via $\xi \sim c^{-\nu_{\perp}}$. Scaling arguments~\cite{Cardy} dictate that the characteristic timescale diverges near $c=0$ as $\tau \sim c^{-\nu_{\parallel}}$, where $\nu_{\parallel} = z \nu_{\perp}$. We plot the logarithm of this timescale, $\log_{10} \tau = \nu_{\parallel} \log_{10} c$, in Figure~\ref{fig:timescale}. We find a similar degree of non-monotonic behaviour (roughly 1 decade at the lowest concentrations $c$) to that shown in Figure 3 of Reference~\cite{PhillnDave}. For consistency we use the exponent $z_3$ calculated to second order. We expect this exponent to be relevant on large wavelengths, for example in equilibrium for $d\gg a$, as was the case for those simulations shown in Figure 3 of Reference~\cite{PhillnDave}.

Lastly, we can infer the lengthscale at which the attractive East model crosses over from near-diffusive to sub-diffusive behavour. To extract such a crossover length, recall that the first order RG result, Equation (\ref{first}), looks like the Liouvillian of an ordinary East model (with rescaled parameters), in that the rates for excitation and relaxation of a cell are unaffected by the state of the neighbour to the east. By contrast, the original Liouvillian (\ref{one}) and the second order RG result (\ref{second}) describe models whose dynamics depends on the state of a cell's east neighbour. 

We thus identify the parameter $\beta$ in Equation~(\ref{params-1}) as a measure of the extent to which the model looks `East-like' (rates insensitive to the state of the east neighbour) or `attractive East-like' (rates dependent upon the state of the east neighbour). We equate the former situation with the regime of hierarchical (dramatically sub-diffusive) coarsening. 

\begin{figure}
\begin{center}
\psfig{file=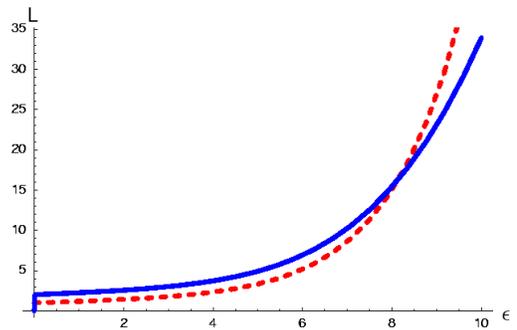,width=2.8in}
\end{center}
\caption{ \label{fig:xo} (Colour online) Crossover length $L_{xo} = (\e_s/\e)^{1/y_{\e}}$ calculated from the second-order RG result, for $c=10^{-3}$, as a function of $\e$  (solid blue line). We show also the phenomenological prediction $a(c=10^{-3},\e)$ (red dashed line). The RG result gives correctly the trend of increasing crossover length with increasing $\e$, as we expect on physical grounds. However, additional physical input is needed to fix the absolute scale of the crossover. Here we obtain close agreement between the two curves by choosing arbitrarily $\e_s = 10^{-3}$, but the RG result varies strongly with $\e_s$ and can be made to look very different from $a(c,\e)$.}

\end{figure}

In the language of RG, then, we expect sub-diffusive behaviour when the renormalized parameter $\e'$ becomes small. Let us define the exponent $y_{\e}$ via $\e' \sim b^{-y_{\e}} \e$. From the linearized second-order result we find $\e' = \e/2$, giving $y_{\e} = 1$. It is more meaningful, however, to retain all orders of $\e$. If we do this, and then iterate the RG until $\e' \sim \e_s$, where $\e_s$ is some sufficiently small value of the renormalized coupling $\e'$, we find the corresponding value of the rescaling parameter: $b \sim (\e_s/\e)^{1/y_{\e}}$. Since this occurs as a consequence of rescaling space by a factor of $b^{-1}$, we infer the crossover length $L_{xo} \sim (\e_s/\e)^{1/y_{\e}}$. On lengthscales $\ell \gg L_{xo}$ we thus expect hierarchical, sub-diffusive relaxation. We plot this crossover length in Figure~\ref{fig:xo}, and find that for $\e_s = 10^{-3}$ we obtain good agreement with the phenomenological crossover length $a(c, \e)$. Note that since $\e_s$ is arbitrary we require additional physical input to fix the absolute scale of the crossover length. The predicted trend of increasing $L_{xo}$ with increasing $\e$, however, is illuminating.

\begin{figure}

\begin{center}
\psfig{file=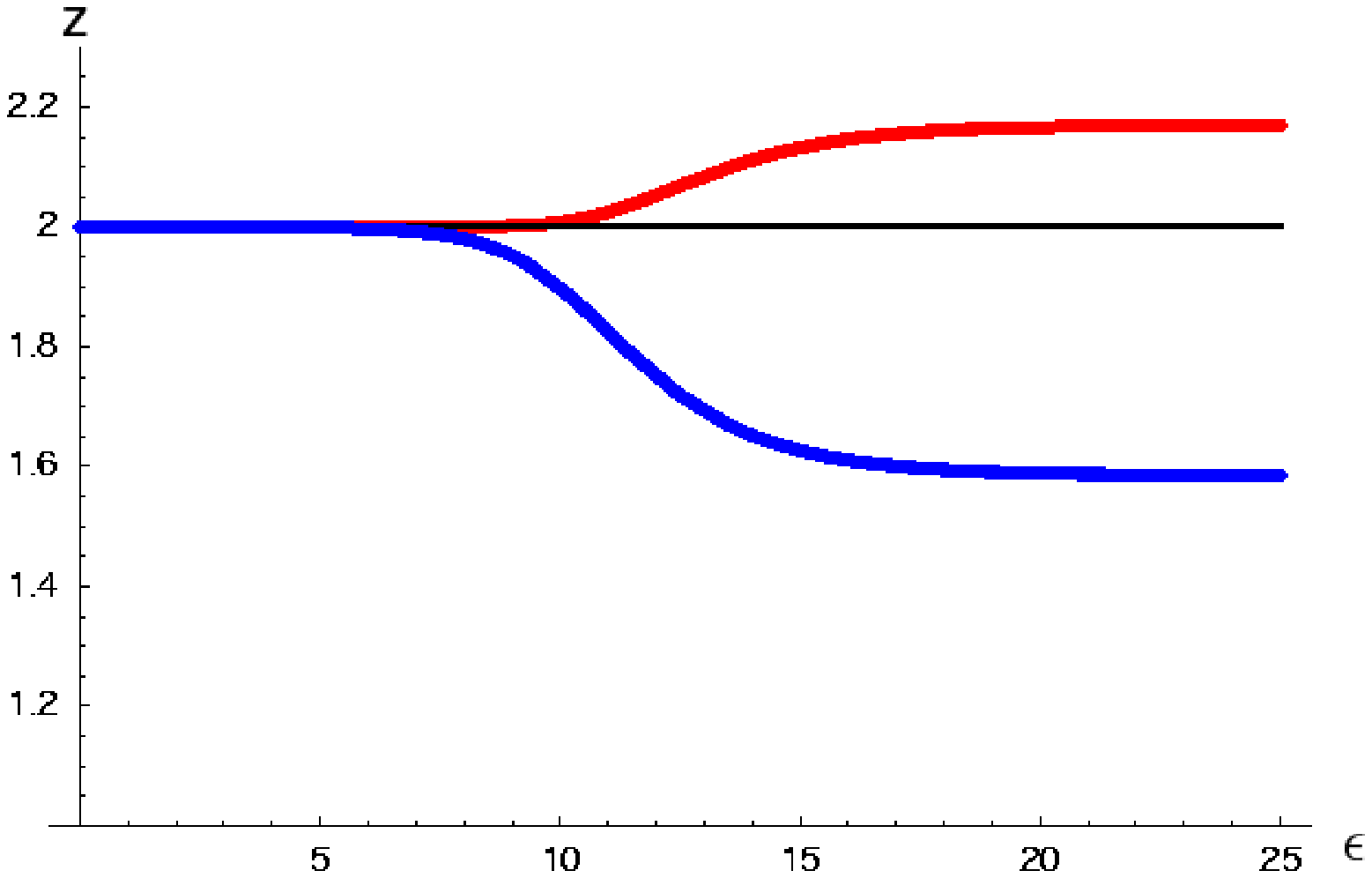,width=2.8in}
\psfig{file=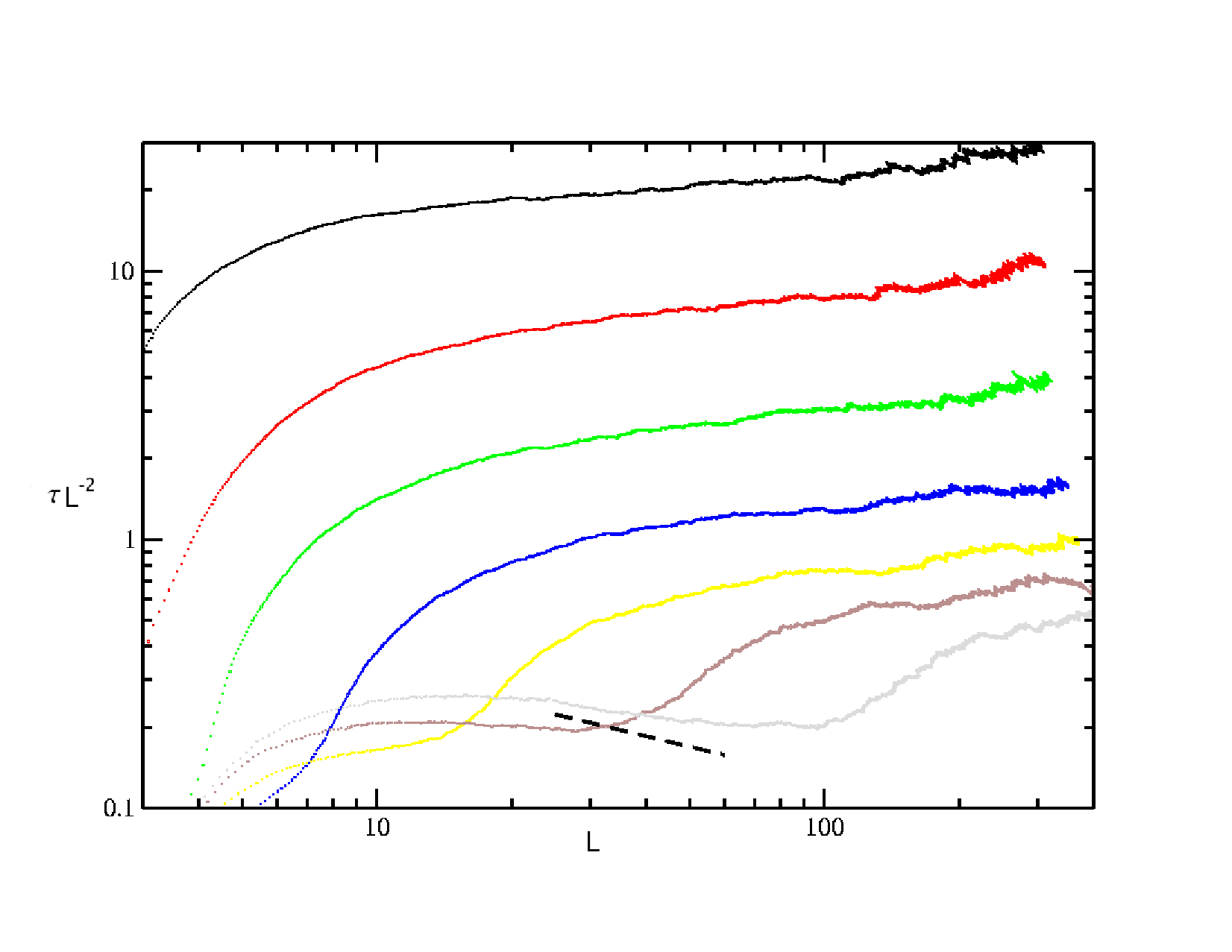,width=2.8in, angle=270}
\end{center}
\caption{\label{fig:fa} (Colour online) Top: Real-space RG prediction for the dynamic exponent $z$ for an attractive FA model as a function of attraction strength $\e$; $c=10^{-3}$. Note the bifurcation for $\e \gtrsim 8$. Bottom: A test of the RG calculation -- Elapsed time rescaled by the square of the mean domain length $\tau L^{-2}$ versus mean domain length $L$ for $c=10^{-3}$ and $\e = 0,2,4,6,8,10,12$ (top to bottom). We show averages over 5 runs for each attraction strength, starting from initial excitation concentration $n(0) = \frac{1}{2}$. For the topmost curves the scaling prediction $\tau \sim L^2$, for which $\tau L^{-2} \sim$ const., is reasonably demonstrated. For $\e > 8$ (the two lowest curves) we see a regime in which $\tau \sim L^{z}$ with $z < 2$, as predicted by the RG calculation. We show as a heavy black dashed line the prediction $\tau \sim L^{1.60}$ for $\e =12$.
}
\end{figure}

As an aside, and a further demonstration of the predictive power of the RG scheme, it is interesting to note that a first-order real-space RG calculation for an attractive version of the FA model suggests a `bifurcation' at large $\e$ of the dynamic exponent (Figure~\ref{fig:fa}, top). Our numerics support this prediction (Figure~\ref{fig:fa}, bottom). Our RG treatment of the attractive FA model is a straightforward adaptation of the procedures used in Reference~\cite{RSRG} and in this paper: we modify the embedding operator to account for the attraction $\e$, and implement the RG to first order.

\section{Conclusions}
\label{sec:conc}
We have predicted and verified numerically that the attractive East model shows a two-stage coarsening behaviour in its out-of-equilibrium regime, as well as a non-monotonic variation of relaxation time with attraction strength. In doing so, we have shown that the real-space RG scheme of References~\cite{Stella, Hooyberghs}, subject to our arbitrary selection of the coarse-graining subspace, and an uncontrolled approximation to second order, captures several interesting characteristics of this model. It also bears out the intuitive idea of Reference~\cite{PhillnDave} concerning the `renormalized East-like' nature of the model at large wavelengths. Because the RG scheme may be motivated in simple physical terms, it is therefore a useful starting point in directing more detailed analyses, either numerical or theoretical.

The two-stage coarsening mechanism of the attractive East model is an example of  two-stage relaxation induced by a competition between a static attraction and a kinetic constraint. It would be interesting to search for such a mechanism in real systems. The attractive East model was intended originally to be studied in its equilibrium regime, at fixed $c$, corresponding to fixed colloid packing fraction. To test the results of this paper against the behaviour of real attractive colloids we propose the following experiment: a sudden pressure increase at fixed $T$, inducing an increase in packing fraction, and hence a reduction in $c$ as the liquid equilibrates. On the basis of our results we would expect particle mobility to differ qualitatively depending on the wavelength one probes and the isothermal compression line one explores~\cite{expts}. For weak attractions and for moderate attractions at large wavelengths we would expect particle motion to be controlled principally by free volume, and thus to be sub-diffusive and repulsive glass-like. For moderate attractions at small wavelengths we would expect particles to move in an approximately diffusive manner, because attractions render the short-wavelength structure of the collid labile.

In general terms, it would be interesting to determine whether two-stage relaxation like that descibed in this paper could be brought about by a competition between two static attractions, one short-ranged and repulsive, and the other long-ranged and attractive~\cite{ref}. 

It would also be interesting to see if the two-time scaling behaviour of the attractive East model could be determined by RG or other means; the different behaviour exhibited by the original East model and its attractive counterpart in the nonequilibrium regime (Figure 3) suggests that the two-time scaling behaviour of the models differ. Since this regime corresponds, according to our mapping, to the aging regime of an attractive colloid, such a study would be a valuable way of comparing the behaviour of the attractive East model and real colloids.

\section{Acknowledgements}
We are grateful to YounJoon Jung, Sander Pronk and Juan P. Garrahan for discussions, and to Carlo Vanderzande for correspondance.

\end{document}